# BLOCKMEDCARE: ADVANCING HEALTHCARE THROUGH BLOCKCHAIN INTEGRATION WITH AI AND IOT

Oliver Simonoski[1], Dijana Capeska Bogatinoska[2]

[1]Faculty of Communication Networks and Security, University of Information Science and Technology, St. Paul the Apostle, Ohrid, 6000, Macedonia
[2]Faculty of Applied IT, Machine Intelligence and Robotics, University of Information Science and Technology, St. Paul the Apostle, Ohrid, 6000, Macedonia

## ABSTRACT

*This research explores the integration of blockchain technology in healthcare, focusing on enhancing the security and efficiency of Electronic Health Record (EHR) management. We propose a novel Ethereum-based system that empowers patients with secure control over their medical data. Our approach addresses key challenges in healthcare blockchain implementation, including scalability, privacy, and regulatory compliance. The system incorporates digital signatures, Role-Based Access Control, and a multi-layered architecture to ensure secure, controlled access. We developed a decentralized application (dApp) with user-friendly interfaces for patients, doctors, and administrators, demonstrating the practical application of our solution. A survey among healthcare professionals and IT experts revealed strong interest in blockchain adoption, while also highlighting concerns about integration costs. The study explores future enhancements, including integration with IoT devices and AI-driven analytics, contributing to the evolution of secure, efficient, and interoperable healthcare systems that leverage cutting-edge technologies for improved patient care.*

## 1. INTRODUCTION

The global healthcare industry faces increasing pressure to achieve higher efficiency and precision in diagnostics, a challenge that digital solutions are addressing. Electronic Health Records (EHR) and Health Information Exchange (HIE) platforms have significantly reduced medical expenses while enhancing healthcare quality [1]. The integration of Internet of Things (IoT) and smart healthcare systems enables remote patient monitoring and communication with healthcare professionals, improving overall well-being [2]. However, the digital transfer of sensitive medical data raises privacy and security concerns, particularly regarding targeted cyber-attacks [3].

Existing healthcare systems often struggle with fragmented data storage, hindering efficient information sharing across providers. This fragmentation impedes interoperability and seamless communication, leading to delays and increased risk of errors in patient treatment. Moreover, centralized systems are vulnerable to data breaches, creating single points of failure exploitable by cyber attackers.

To address these persistent issues, the healthcare sector is turning to blockchain technology. As a decentralized and immutable ledger, blockchain is reshaping various industries, including finance, logistics, and healthcare [4]. Its impact extends beyond economic benefits to political, humanitarian, social, and scientific domains, offering potential for improved governance and enhanced collaboration.

Our blockchain-integrated solution tackles the problems of fragmented data storage and security vulnerabilities in healthcare systems. It represents a significant advancement over traditional approaches by creating a unified framework that ensures both data security and seamless sharing among stakeholders [5]. Leveraging blockchain's decentralized architecture, we create a comprehensive system that secures patient data while facilitating real-time access for authorized providers. The consensus mechanism prevents unauthorized alterations, ensuring data integrity [6], while cryptographic hashing adds an additional layer of protection.

Blockchain technology also enhances transparency and audit capabilities [7]. Each transaction is logged with a timestamp and unique cryptographic identifier, providing a clear history of data modifications. This feature is particularly valuable in healthcare, where audit trails are essential for compliance and accountability. Our approach also empowers patients with greater control over their information access, fostering trust in the healthcare process.

This paper builds upon and significantly extends our previous work presented at the CCSITA 2024 conference and published in the International Journal on Cybernetics & Informatics (IJCI) [8]. In our initial study, 'BlockMedCare: Advancing Healthcare Through Blockchain Integration,' we introduced the concept of using blockchain technology to enhance healthcare data management. The current paper expands on this foundation by incorporating more advanced features, including the integration of AI and IoT technologies, a more comprehensive user survey, and an in-depth analysis of ethical and societal implications. We also provide a more detailed examination of the system's architecture and implementation, offering insights into practical deployment challenges and solutions.

Unlike existing blockchain healthcare solutions that primarily focus on data security, our proposed framework uniquely integrates patient-centric care, optimized diagnostic processes, and enhanced data security within a single, comprehensive system. This holistic approach addresses current healthcare challenges while laying the groundwork for a more interconnected and efficient healthcare ecosystem.

Our research aims to overcome barriers to blockchain adoption in healthcare by proposing a framework that enhances patient-centric care, optimizes diagnostic accuracy, and strengthens data security. By empowering patients and fostering trust among stakeholders, we aim to create a more resilient and efficient healthcare infrastructure.

This paper is organized as follows: We begin with an exploration of blockchain technology's theoretical foundations, followed by our Healthcare Development Methodology. We then detail the System Design and Architecture, Implementation and Testing processes, and present a critical evaluation in our Results and Data section. Finally, we synthesize our findings, offering insights into future work and recommendations for further system enhancement.

## 2. RELATED WORK AND CURRENT LANDSCAPE

The field of medical records management has evolved significantly, transitioning from manual processes to digital solutions. However, both traditional and early digital methods face challenges in ensuring data security, accessibility, and interoperability. This section examines recent advancements and persistent limitations in electronic health record (EHR) management systems. Historically, healthcare institutions have relied on manual processes or web-based data management systems, both presenting significant vulnerabilities. These methods, while aiming to make patient information accessible and organized, face limitations in data handling and are susceptible to breaches. The authors of [1] note that the lack of a standardized, innovative framework has led to disparate technologies struggling with interoperability. Furthermore, a case study in [3] emphasizes that despite system-based advancements, there remains no consensus on the technical infrastructure required for effective patient-centered interoperability.

Recent attempts to address these challenges have focused on two main technological strategies:
*Cloud-based Solutions*:

Researchers [9] proposed a system architecture for managing EHRs using cloud services with detailed access control mechanisms. This intercloud setup integrates multiple cloud environments, offering comprehensive privacy protection and simplified data migration. However, it remains vulnerable to exploits that could compromise patient data privacy.

*Blockchain-based Solutions*:

1. The Action-EHR framework [10], built on Hyperledger Fabric, employs compliant cloud storage and advanced encryption methods. Despite strengths in authentication and authorization, it faces challenges in key management and requires multiple uploads of patient records for different providers.
2. A hybrid blockchain model [11] aims to balance security and privacy with high availability and efficiency. However, it faces limitations in real-world testing and potential privacy issues related to public blockchain segments.

Our proposed system addresses several critical gaps in existing approaches:

1. *Comprehensive Integration*: Unlike studies focusing on specific aspects, our research proposes a complete EHR system integrating multiple healthcare functions within a single, blockchain-based framework.
2. *Enhanced Security and Usability*: We incorporate role-based access control and a user-friendly interface, addressing both security and usability concerns often overlooked in previous studies.
3. *Scalability and Efficiency*: Our system addresses common limitations such as scalability issues and high computational costs, leveraging blockchain's decentralized nature to enhance data security and privacy while maintaining real-world flexibility.
4. *Simplified Key Management*: We implement a streamlined process that reduces complexity for end-users, improving accessibility for both patients and healthcare providers.
5. *User-Centric Approach*: We include a user survey to gain insights into potential adoption challenges, focusing on practical aspects such as user training, regulatory compliance, and real-world applicability.

Figure 1 summarizes the key differences between our blockchain-integrated approach and traditional healthcare data management systems, highlighting improvements in security, interoperability, and patient empowerment.

## 2.1. Statistics on Blockchain Adoption in Healthcare

The global healthcare industry is gradually adopting blockchain technology to address data security, patient privacy, and interoperability issues. Recent studies project significant growth in the blockchain healthcare market, driven by increasing awareness of data breaches and the need for secure patient data management [12].

*Potential Use Cases*

- *Supply Chain Management*: Tracking pharmaceuticals and medical supplies to ensure authenticity and prevent fraud.
- *Clinical Trials and Research*: Enhancing transparency and security in data storage and sharing.
- *Patient Consent Management*: Allowing patients to control access to their health records with immutable consent verification.

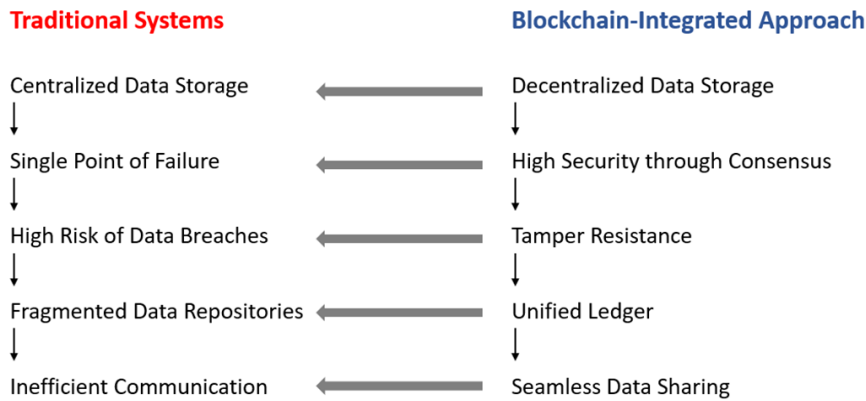

Figure 1. Traditional systems vs blockchain-integrated approach

The global blockchain healthcare market, valued at $0.76 billion in 2022, is projected to reach $14.25 billion by 2032, representing a compound annual growth rate (CAGR) of 34.02% from 2023 to 2032, as illustrated in Figure 2 [11].

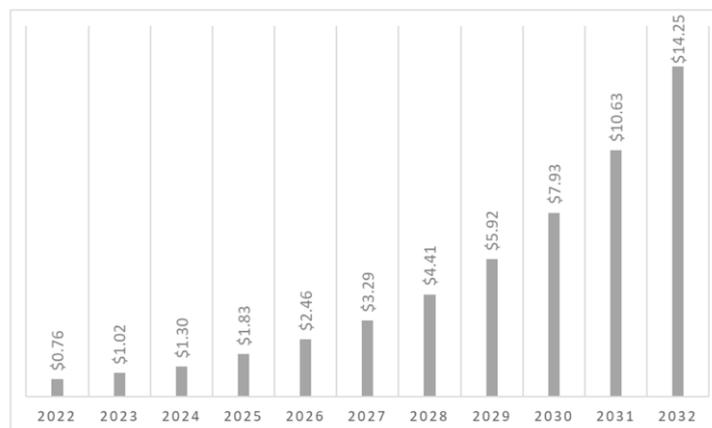

Figure 2. Projected Growth of the Global Blockchain Healthcare Market (2022–2032)
Source: Blockchain in Healthcare Market Report [11]

In summary, our approach provides a more integrated and user-centric solution compared to existing systems. By addressing limitations in scalability, key management, and practical implementation, our research contributes significantly to creating a more secure, efficient, and patient-centric healthcare infrastructure.

# 3. THEORETICAL FOUNDATIONS OF BLOCKCHAIN IN HEALTHCARE SYSTEMS

Blockchain technology has emerged as a powerful tool for enhancing data security, transparency, and decentralization across various industries, including healthcare. This section provides a comprehensive review of blockchain's fundamental principles, evolution, and specific applications in the healthcare sector.

Blockchain is a distributed ledger technology that records transactions across a network of nodes, offering a decentralized approach that enhances security and reliability. Unlike centralized systems vulnerable to single points of failure, blockchain coordinates independent nodes collaboratively, ensuring data integrity and availability without a centralized supervisor [13].

In healthcare, a distributed model effectively balances security and accessibility, protecting patient data while allowing continuous access and collaboration among providers. Blockchain has evolved from its initial use in cryptocurrency (blockchain 1.0) to smart contracts (blockchain 2.0), and now to healthcare applications with blockchain 3.0. This latest iteration focuses on scalability, interoperability, and environmental sustainability, featuring advanced consensus algorithms like proof of stake and sharding. In healthcare, blockchain 3.0 offers significant potential for improving patient data management, enhancing interoperability, and ensuring medical record integrity and security.

Figure 3 illustrates the operational stages of blockchain, including transaction handling, block formation, and data synchronization across the network.

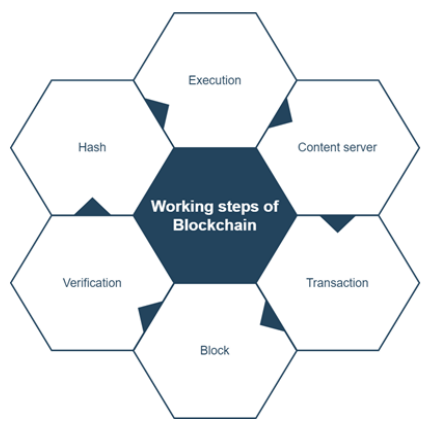

Figure 3. Operational steps of blockchain technology

Blockchain's peer-to-peer data transfer reduces costs and ensures confidentiality compared to traditional e-health systems relying on third-party intermediaries. It employs asymmetric encryption methods, such as elliptic curve cryptography and RSA signature algorithms, for secure communication. The Secure Hash Algorithm (SHA) plays a crucial role in generating unique block fingerprints, ensuring data integrity and preventing unauthorized modifications [14].

Key features of blockchain technology include:

1. *Security*: Maintained through cryptographic techniques
2. *Transparency*: Achieved through publicly accessible ledgers
3. *Participation*: Ensured by involving multiple nodes in the validation process

Key components include platforms like Ethereum, smart contracts, mining processes, and consensus mechanisms.

## 3.1. Classification of Blockchain and Consensus Mechanism

Blockchain systems are classified into four categories based on node selection approaches: public, consortium, private, and community blockchains [15]. Each type offers different levels of security and decentralization, suitable for various use cases. In healthcare, the choice depends on factors such as data privacy requirements, regulatory compliance, and interoperability needs. Consensus mechanisms are crucial for synchronizing nodes and ensuring agreement on legitimate transactions. Common algorithms include:

- *Proof of Work (PoW)*: Secure but resource-intensive and time-consuming;
- *Proof of Stake (PoS)*: More energy-efficient, based on ownership stake;
- *Delegated Proof of Stake (DPoS)*: Faster transactions through elected trustees;
- *Other Consensus Mechanisms*: Proof of Capacity (PoC) and Proof of Elapsed Time (PoET).

Table 1 summarizes key properties of consensus mechanisms, reflecting real-time data from the research and analysis stages.

Table 1. Comparison of Consensus Mechanisms

| Consensus mechanism | PoW | PoS | DPoS |
|---|---|---|---|
| Scenes | Public chain | Public chain Alliance chain | Public chain Alliance chain |
| Accounting nodes | All nodes | All nodes | Select representative nodes |
| Response time | ~ 10 minutes | ~ 1 minute | <1 minute |
| Ideal state Of Transaction Per Second | 7 TPS | 300 TPS | 500 TPS |
| Fault tolerance | 50% | 50% | 50% |

## 3.2. Blockchain Architecture and its Implications

Blockchain architecture comprises five core layers [16]:

- *Application Layer*: Encodes end-user functionalities;
- *Execution Layer*: Handles instructions and smart contract execution;
- *Semantic Layer*: Validates transactions and defines data structures;
- *Propagation Layer*: Manages peer discovery and message transmission;
- *Consensus Layer*: Ensures agreement on ledger state.

Blockchain's core properties - immutability, decentralization, and verifiable information validity make it well-suited for applications requiring secure, transparent, and tamper-proof data management. Its ability to facilitate cross-border transactions and maintain permanent records underscores its potential in scenarios involving multiple untrusted participants.

In conclusion, blockchain technology, especially in its latest iteration (blockchain 3.0), presents innovative solutions for managing healthcare data. Its secure, transparent, and decentralized architecture could potentially transform healthcare delivery, optimize patient outcomes, and increase overall healthcare service effectiveness.

# 4. ARCHITECTING A BLOCKCHAIN-BASED HEALTHCARE SOLUTION: FRAMEWORK AND METHODOLOGY

This section outlines the methodological framework and approach employed in developing our blockchain-based healthcare solution. We describe the three-layer architecture forming the foundation of our system, examine functional and non-functional requirements, and explain our research methodology.

## 4.1. System Architecture

Our healthcare platform is organized into three layers, chosen for their ability to separate concerns, enhance modularity, and improve scalability:

- *User Interface Layer*: Responsible for displaying data and receiving user input, with interfaces compatible with both mobile and desktop devices;
- *Business Logic Layer*: Facilitates interaction between users and the blockchain-based EHR platform, ensuring consistent data interfaces and standards. It encapsulates user data into virtual transactions and resources, transferring data to blockchain nodes and storing new user data within the blockchain database;
- *Data Access Layer*: Manages transaction verification, block creation, and consensus processes. It updates the blockchain database and maintains a customized blockchain containing all verified EHRs.

This architecture creates a resilient, integrated solution enhancing system efficiency and data management while ensuring broad accessibility.

## 4.2. System Requirements

Expanding on this architectural foundation, we outlined several crucial requirements necessary for the creation of a reliable and secure electronic health records platform. These requirements were developed through the design science process described in the preceding section.

### 4.2.1. Functional Requirements

1. Establish a registration process for all users;
2. Enable patient control over their records;
3. Permit doctors to submit data into the system;
4. Ensure limited access for authorized entities to download information;
5. Apply encryption to all records and stored information.

Additionally, the system mandates that all users access their accounts through MetaMask, with authentication required for login. Patients can view their medical records anytime and anywhere, as long as they have a connection to the blockchain. The system supports appointment management, allowing patients to request appointments and doctors to retrieve these requests, update appointment details, and prescribe medication. Doctors can also input and send laboratory results directly to patients, ensuring a streamlined process for managing health information. All users have the capability to log out of the system, maintaining control over their sessions.

### 4.2.2. Non-Functional Requirements

- *Accessibility*: Compatible with any blockchain-connected browser;
- *Usability*: Intuitive design for users of all technological backgrounds;
- *Security*: Unique transaction hash for each request;
- *Data Integrity*: Validation of all entered data.

### 4.3. Research Methodology

We use an experimental design to systematically evaluate previous achievements and identify similar solutions. This approach allows for controlled testing and detailed analysis, which is essential given the limited number of prior implementations that meet our system's requirements. Our object-oriented analysis methodology incorporates use-case diagrams and sequence modeling to elucidate functional needs and represent system processes. This introduces a patient control system that consolidates all health data into a single, accessible platform.

Figure 4 illustrates the EHR System Architecture, depicting how data moves through each layer of our system. It highlights key processes such as user authentication, record access, and transaction verification, providing a comprehensive view of the system's operation.

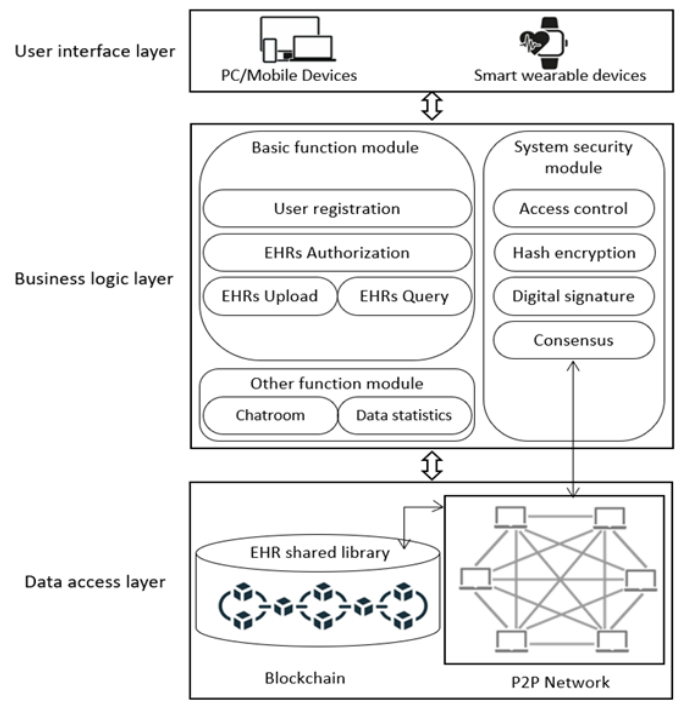

Figure 4. EHR System Architecture

Our platform provides comprehensive access to medical history and examinations for authorized entities, streamlining data access for healthcare providers and patients. By centralizing health data management, we address the growing demand for patient empowerment and data ownership, enhancing patient control over health records while improving overall system efficiency and accessibility.

In conclusion, our methodological framework—integrating a three-layer architecture, comprehensive system requirements, and an experimental research design—establishes a solid foundation for developing a blockchain-based healthcare solution that addresses key challenges in electronic health record management.

# 5. INTEGRATING BLOCKCHAIN, AI, AND IOT FOR ADVANCED HEALTHCARE MANAGEMENT

This section outlines the system design and architecture of our blockchain-based healthcare solution, describing key entities, blockchain infrastructure, smart contract implementation, and role-based access control (RBAC) mechanisms.

## 5.1. System Overview and Key Entities

Our healthcare system is built on a blockchain framework with three main entities: patients, doctors, and administrators. The system uses cryptography to protect medical data, granting access only with patient consent. Figure 4 illustrates the entities, workflow, and operations within the system.

### 5.1.1. Entity Roles

*Administrator Node:*

The administrator node serves as the central authority, handling smart contract deployment, user authentication, medication management, and system settings. It also facilitates data export and manages lab test parameters, ensuring smooth system operations. Other entities rely on it for essential authentication and configuration.

*Patient Node:*

As shown in Figure 5, the patient node manages personal identification information and the patient's Ethereum address. Patients can update personal details and schedule appointments with specific doctors.

*Doctor Node:*

While not explicitly detailed in the provided text, we can infer that doctor nodes have functionalities related to accessing patient records, managing appointments, and inputting medical data.

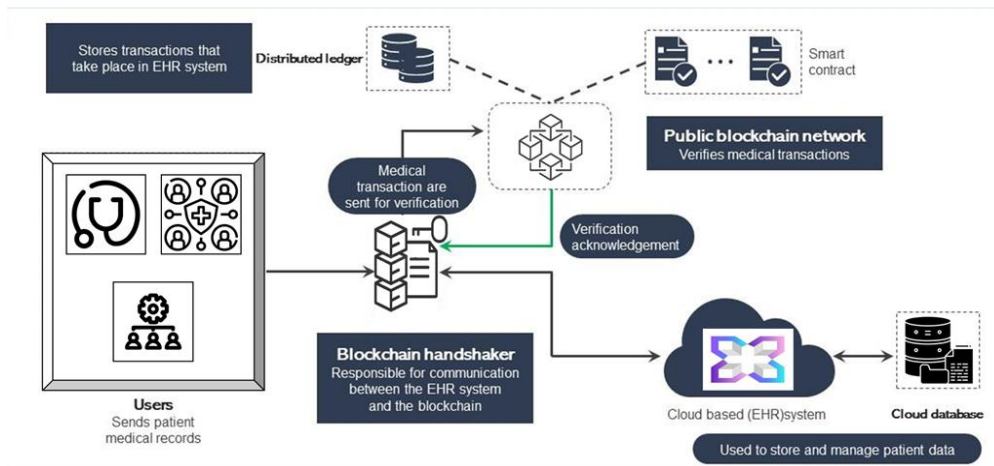

Figure 5. Architecture of proposed system model

## 5.2. Blockchain Infrastructure

The system uses a peer-to-peer (P2P) network for efficient EHR data replication. This decentralized setup enhances data availability, reduces single points of failure, and strengthens system resilience.

## 5.3. Smart Contracts

Smart contracts enable secure and automated interactions among entities, imposing strict access controls and regulating user permissions. They streamline various operations including user login/registration, data management, appointment scheduling, and data exportation.

## 5.4. Role-Based Access Control (RBAC)

The EHR system employs RBAC with three default roles: patients, doctors, and system administrators. Each role has specific permissions:

- *Patients:* Perform actions related to their individual EHRs;
- *Doctors:* Add, query, and analyze EHRs, and engage in information exchange;
- *System administrators:* Modify user statuses, set system start dates, and adjust overall settings.

Figure 6.a visually represents the RBAC structure, illustrating how the system determines user access based on their role.

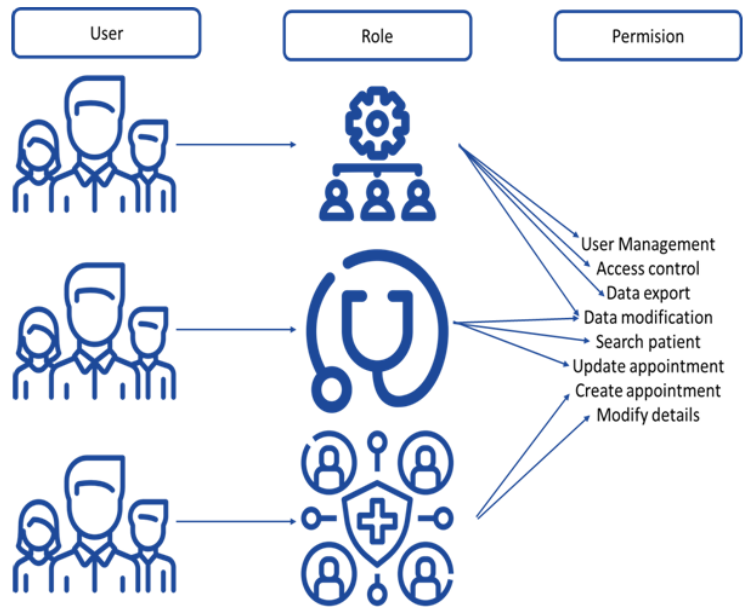

Figure 6. a) Role based EHR structure

The RBAC implementation follows the algorithm represented in Figure 6.b:

1. Check if user is signed in.
2. Prompt for login if not signed in.
3. Retrieve user ID and validate permissions based on role.
4. Obtain role from smart contract using ID.
5. Grant or deny access based on operation permission.

## 5.5. AI Integration in Healthcare Systems

The integration of Artificial Intelligence (AI) and blockchain technology is positioned to fundamentally reshape healthcare systems. Blockchain ensures secure data management, protecting patient privacy, while AI complements this by providing powerful data analytics, predictive modelling, and personalized treatment recommendations [17]. Together, these technologies enable healthcare systems to process, analyse, and utilize data more effectively. The AI-driven insights enhance the decision-making capabilities of healthcare providers, leading to improved patient outcomes and optimized care processes. Within the system architecture, AI algorithms can be employed in the ***Business Logic Layer***, where patient data stored on the blockchain is analysed to produce real-time insights. AI models can forecast disease outbreaks, identify high-risk patients, and even automate diagnostic processes, reducing human error [18].

*AI-Driven Insights for Enhanced Diagnostics:*

AI-driven predictive analytics can further empower healthcare providers by leveraging decentralized healthcare data stored on the blockchain. These models can automate diagnostic processes and provide early warnings about potential health issues, which can improve patient outcomes by reducing diagnostic errors. Moreover, AI can analyse large-scale anonymized health datasets stored on the blockchain, offering insights into patient histories while adhering to strict privacy standards enforced by blockchain's access control mechanisms.

Incorporating machine learning algorithms within this architecture enables hospitals and healthcare systems to gain real-time insights, such as early detection of disease patterns. This not only improves efficiency but also minimizes delays in treatment [19]. Blockchain ensures that sensitive patient data remains secure throughout the AI-driven analysis process, allowing only authorized personnel to access or alter the information.

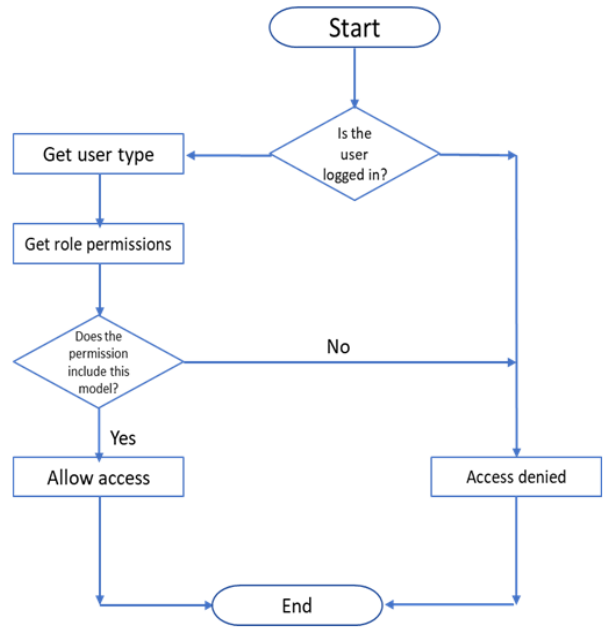

Figure 6. b) RBAC structure algorithm

## 5.6. Interoperability with IoT Devices in Healthcare

The proliferation of Internet of Things (IoT) devices in healthcare has transformed the way patient data is collected, monitored, and used in treatment. Devices like wearable fitness trackers, continuous glucose monitors, and remote monitoring systems for chronic disease management allow for continuous data collection, improving patient outcomes and enabling doctors to receive real-time insights into patient health. However, despite these benefits, the secure sharing of IoT-generated healthcare data remains a challenge, as many systems lack the appropriate data-sharing frameworks [20].

*Leveraging Blockchain for IoT Data Security*

Blockchain technology can address these challenges by providing a decentralized, secure platform for managing IoT-generated healthcare data. By storing data from IoT devices on the blockchain, healthcare providers can ensure that patient information is secure, tamper-proof, and accessible only to authorized personnel.

Moreover, the decentralized nature of blockchain facilitates efficient real-time data sharing across multiple healthcare providers, enhancing care coordination. For instance, a patient wearing a smart glucose monitor could securely share real-time data with their healthcare provider via blockchain, enabling timely intervention when necessary [21].

In conclusion, our blockchain-based healthcare solution integrates advanced security measures, efficient data management, and user-friendly interfaces. The combination of blockchain technology, smart contracts, RBAC, AI integration, and IoT interoperability creates a robust framework for managing electronic health records while ensuring data privacy and integrity.

# 6. IMPLEMENTATION AND EVALUATION OF A BLOCKCHAIN-BASED HEALTHCARE SYSTEM

In the dynamic realm of healthcare, the Electronic Health Records (EHR) system has undergone thorough validation procedures to ensure accuracy, reliability, and compliance with industry standards [22]. Despite these validation efforts, the need for strong security and data integrity persists. In this context, blockchain technology emerges as a transformative solution, providing an immutable ledger for all transactions within the EHR system. By utilizing blockchain technology, patient records are safeguarded against tampering and unauthorized access, thereby enhancing the security and trustworthiness of healthcare data.

This chapter details the design, implementation, and evaluation of our blockchain-based healthcare solution, focusing on the development process, user interface functionality, and user feedback.

## 6.1. Designing and Deploying Blockchain Solutions for Healthcare

### 6.1.1. Technology Stack and Development Process

Our system utilizes blockchain technology to enhance security, privacy, and integration in healthcare data management. The decentralized application (dApp) is powered by Node.js and React.js, chosen for their robust performance and compatibility with blockchain development (Figure 7).

The development process involves:

- Setting up the production environment (back-end and front-end);
- Integrating smart contracts written in Solidity;
- Testing and deployment using Truffle suite and Ganache;
- Integration of MetaMask for secure transactions.

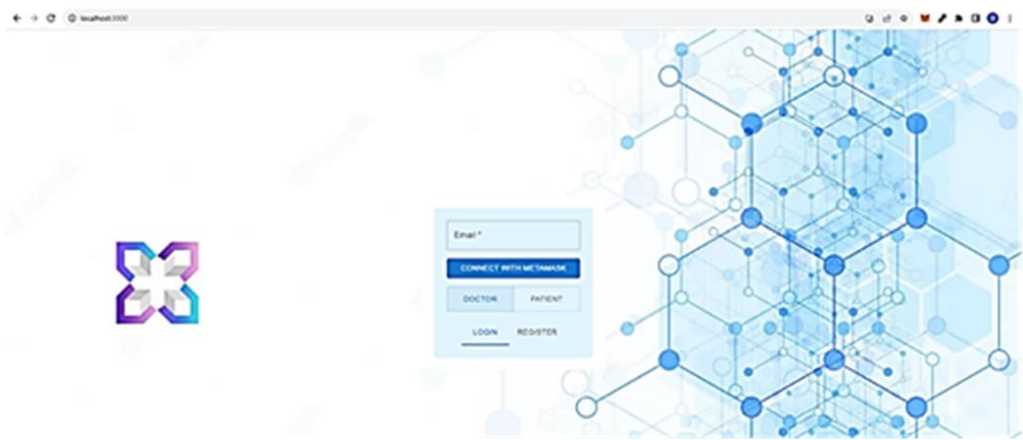

Figure 7. MedCare login/registration page

**6.1.2. User Interface Design and Functionality**

The system's interface prioritizes user convenience across different roles, as demonstrated in successful blockchain implementations in healthcare. For instance, a notable application of blockchain technology in healthcare involved its deployment within a major hospital network to manage patient consent for data sharing among various healthcare providers. This implementation not only enhanced patient control over data access but also led to a marked decrease in disputes related to consent and streamlined administrative processes within the hospital network [23].

*Doctors' Interface:*

The interface for doctors offers intuitive double-click interactions and a personalized dashboard for efficient appointment management (see Figures 8.a and 8.b). Key features include:

- Intuitive appointment management dashboard;
- Specialized tab for laboratory test results;
- "E-reports" tab providing a comprehensive history of past activities

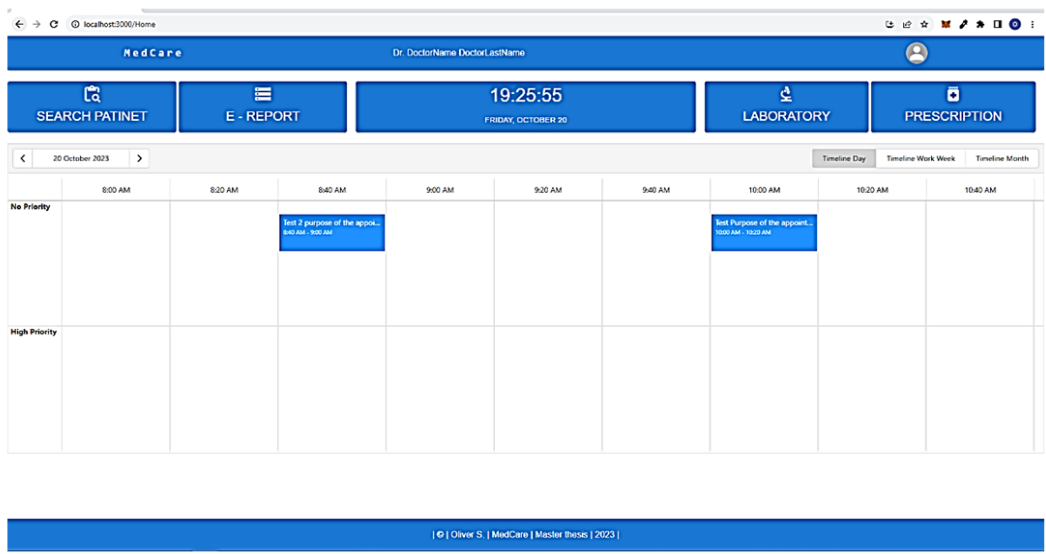

Figure 8.a) MedCare: Doctor dashboard

*Patients' Interface:*

The patient module emphasizes ease of use, transparency, and data integrity. Key features include:

- Easy appointment scheduling (Figure 9);
- Medical history tracking;
- Structured presentation of laboratory results.

*Administrators' Interface:*

Administrators have access to specialized panels for comprehensive data management and oversight. Key features include:
- User management and activation;

- Medication inventory management;
- Laboratory test parameter configuration;
- Data export capabilities in various formats (CSV, XML, TXT) for external analysis (Figures 10 and 11).
-

Figure 8.b) MedCare: Doctor appointment details

Figure 9. MedCare: Patient appointment scheduling

Figure 10. MedCare: Admin data administration pag

Figure 11. MedCare: Admin laboratory administration page

### 6.1.3. Ethical and Societal Considerations: Empowering Patients vs. Responsibility

While implementing blockchain technology in healthcare offers numerous benefits, it also raises critical ethical concerns. The decentralization of healthcare data through blockchain empowers patients by giving them control over their health information, allowing them to decide who can access it and under what circumstances. This shift toward patient autonomy is significant, yet it brings with it the added responsibility of making informed decisions about data sharing, particularly in complex situations like medical research or clinical trials.

Ensuring that patients are well-informed about the risks and benefits of sharing their health data is essential. Ethical frameworks need to be developed to guide consent processes, ensuring that patients fully understand the implications of their choices, especially when blockchain is used to manage sensitive health data. Additionally, the potential risks of technology misuse, such as unauthorized access or exploitation of data, demand robust governance structures. By addressing these ethical concerns, blockchain implementation can align with the broader goal of promoting patient-centered care while safeguarding individual rights.

*Societal Impact: Bridging the Healthcare Gap*

Blockchain technology has the potential to democratize access to healthcare services, offering secure, decentralized solutions that can empower patients in remote or underserved areas. However, this promise is tempered by the risk of exacerbating the digital divide. Populations lacking access to the necessary technology may be left behind, further deepening healthcare disparities, particularly among socioeconomically disadvantaged communities. On the other hand, by decentralizing health data and facilitating telemedicine platforms, blockchain can help bridge these gaps by connecting patients in underserved regions with world-class medical expertise. To maximize these benefits, governments and policymakers must ensure that blockchain-based healthcare services are accessible and affordable to all, preventing new barriers to care from emerging. A balanced approach to implementing these technologies is critical to ensuring that they truly enhance healthcare access rather than reinforcing existing inequalities.

## 6.2. Testing and survey analysis: Ensuring reliability and exploring user perspectives

Having detailed the design and implementation of our blockchain solution, we now turn to the crucial aspects of testing and user feedback to ensure the system's reliability and usability.

### 6.2.1. Testing Methodology

The validation process ensures real-time accuracy in patient record updates, supporting continuous care. Testing methods play a crucial role in ensuring the reliability and functionality of the system, incorporating:

- Usability and functionality testing;
- Measurement of response times and error rates;
- Use of Ethereum Virtual Machine (EVM) and Ganache for local blockchain development.

### 6.2.2. Survey Methodology and Results

To gain deeper insights into the usability and user-friendliness of the design, we conducted a survey targeting different groups of participants, including IT professionals and healthcare workers. The survey consisted of 28 questions covering various aspects related to design usability and user experience. Participants were selected to represent a diverse range of ages and occupations within the healthcare and IT sectors.

*Survey Results*:

- 66% of respondents reported moderate familiarity with blockchain technology, with only 13% indicating no familiarity at all (Figure 12.a).
- Responses regarding prior EHR system usage were evenly split, with 15 affirmatives and 15 negatives (Figure 12.b).
- Only 20% of respondents expressed concerns about the performance and scalability of blockchain technology in larger-scale applications (Figure 13.a).
- Most respondents (93.4%) believe in blockchain's potential to enhance interoperability and sharing across different sectors (Figure 13.b).

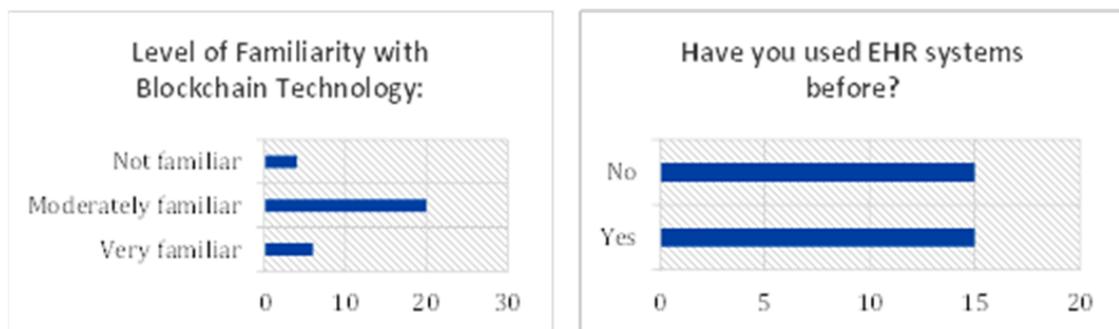

Figure 12.a. Results of a survey / Figure 12.b. Results of a survey

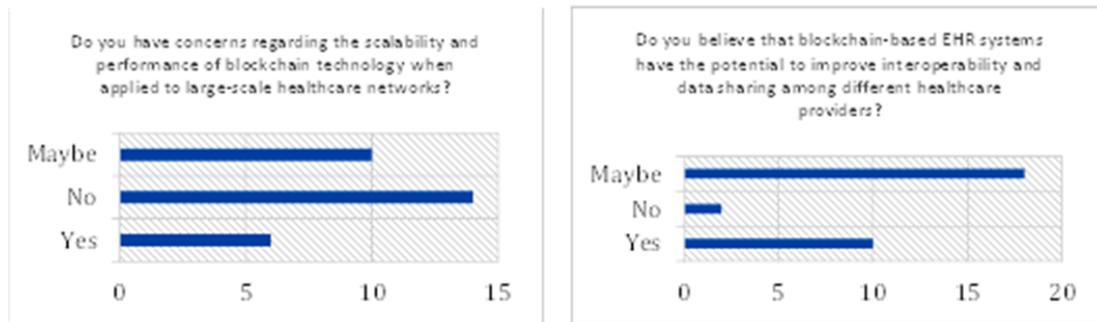

Figure 13. a): Participants' perspectives on the performance and scalability
b): Participants' perspectives on the potential of blockchain technology

*Key Findings*:

- Respondents noted several advantages over current systems, particularly in terms of data sharing among providers and paperwork reduction.
- They advocated for including blockchain education in courses while expressing concerns about integration costs and perceived technology complexity.
- Participants highlighted the critical roles of IoT and AI in integration, underscoring their importance in development and analytics.
- Overall, participants expressed strong interest in future discussions and research in this area.

## 6.3. Security and Privacy

*Privacy and Compliance with Data Regulations*

In the current era of stringent data privacy regulations, blockchain technology offers a robust solution for healthcare providers to ensure compliance while maintaining transparency. Healthcare organizations must adhere to laws such as the **Health Insurance Portability and Accountability Act** (HIPAA) and the **General Data Protection Regulation** (GDPR) [24]. Blockchain's decentralized, cryptographic nature can enforce these privacy standards, ensuring that sensitive patient data is stored securely and accessed only through verified digital signatures. By decentralizing patient data, blockchain eliminates the risks associated with centralized data storage systems, which are prone to breaches. Blockchain's cryptographic techniques ensure that only authorized individuals can access or modify patient records, aligning healthcare operations with global privacy regulations.

*Ensuring Regulatory Compliance with Blockchain*

Blockchain systems provide an immutable and auditable ledger of all transactions, making it easier for healthcare organizations to track who has accessed or modified patient data—a critical requirement for demonstrating compliance with privacy regulations. To further enhance privacy, blockchain systems can implement zero-knowledge proofs, which allow healthcare providers to verify patient data without revealing the data itself. This ensures that patient confidentiality is maintained while healthcare professionals can still access the necessary information for informed decision-making [24]. Moreover, the auditable nature of blockchain enables healthcare institutions to easily demonstrate regulatory compliance during audits, providing an extra layer of security.

In conclusion, our blockchain-based healthcare solution shows promising potential for enhancing data security, patient control, and overall system efficiency. Future work should focus on addressing integration costs and education to facilitate wider adoption of this innovative approach to healthcare data management.

# 7. CONCLUSION AND FUTURE DIRECTIONS

This research investigates the integration of blockchain technology in healthcare, with a specific focus on enhancing the security and efficiency of data sharing. Our study addresses critical challenges associated with electronic health records, including confidentiality, scalability, user privacy, and the implementation of user-friendly cryptographic measures.

The primary focus of our research has been the integration of blockchain into healthcare systems to bolster security protocols, particularly for patient and doctor entities. Our proposed system emphasizes security through patient consent, authentication at each stage, and the use of public key signatures for data transactions, all validated by the blockchain. By leveraging cryptographic mechanisms and smart contract codes, we have created a secure environment that aligns with the goal of developing a security-centric healthcare system.

Survey results indicate a strong interest in adopting blockchain technology among participants, with many recognizing its potential to enhance data sharing and privacy. However, concerns were also raised regarding security, cost, and integration within existing healthcare infrastructures. Despite these challenges, the potential benefits of blockchain in healthcare are substantial. The proposed method's limitations, such as scalability and integration with current systems, suggest that a combined approach utilizing both traditional and blockchain technologies may be a practical first step, allowing for gradual adaptation and testing of blockchain's advantages.

For successful real-world deployment, several practical aspects must be addressed. These include ensuring regulatory compliance with standards such as HIPAA or GDPR, effectively training healthcare professionals, and overcoming adoption barriers through pilot programs and collaborative efforts with industry stakeholders. Additionally, exploring blockchain 3.0 technologies and managing integration costs through a hybrid model approach will be crucial for scalability and feasibility.

Looking forward, future research should focus on a range of critical areas. Conducting experimental security tests will be essential to evaluate the system's resilience against potential attacks. Exploring migration to blockchain 3.0 technology and optimizing blockchain to address its inherent limitations will improve performance and scalability. Enhancing the application's functionality and user interface will ensure accessibility and ease of use.

Furthermore, integrating Internet of Things (IoT) devices for real-time data monitoring and analysis, and incorporating artificial intelligence (AI) to facilitate predictive analytics and personalized medicine, represent exciting avenues for advancement. Consolidating foundational components and offering succinct health data summaries through analytical tools will improve data utility. Exploring health data forensics and establishing a messaging system aligned with health data exchange standards will further enhance security and interoperability.

Government-led initiatives could play a crucial role in facilitating the adoption of blockchain technology in healthcare. By establishing standardized frameworks and policies that promote interoperability and data sharing across different healthcare systems, such initiatives can help create a unified national or regional healthcare network, reducing data duplication and enhancing the continuity of care.

In conclusion, while challenges remain, the integration of blockchain technology in healthcare offers promising solutions to longstanding issues of data security, privacy, and interoperability. By addressing the identified challenges and pursuing the outlined future directions, blockchain has the potential to revolutionize healthcare delivery, significantly improving patient outcomes and system efficiency. As we continue to refine and expand this technology, we move closer to a more secure, efficient, and patient-centric healthcare ecosystem.

## AUTHORS

**Oliver Simonoski** received a Master's degree in Science in Information Technology from the University of Information Science and Technology "St. Paul the Apostle" in Ohrid, Macedonia. As a software developer, he focuses on innovation and efficiency in technology solutions, consistently integrating advanced approaches to enhance system performance and user experience. His main research interests include optimization and distributed computing, cloud technologies, IoT, blockchain technologies, and artificial intelligence.

**Dr. Dijana Capeska Bogatinoska** is an associate professor and researcher at the University of Information Sciences and Technologies "St. Apostle Paul" in Ohrid. With a background in electrical engineering and a doctorate in Applied Computer Science, she has a strong expertise in the field. Dr. Capeska Bogatinoska has participated in various national and international projects, including Horizon 2020, Erasmus+, and COST Actions. She has published over 60 papers in scientific conferences and journals and serves as a reviewer for reputable publications. Additionally, she is actively involved as a TCP member in multiple conferences and holds a position in the university senate.